\documentclass[conference]{IEEEtran}
\IEEEoverridecommandlockouts

\usepackage{cite}
\usepackage{amsmath,amssymb,amsfonts}
\usepackage{algorithmic}
\usepackage{graphicx}
\usepackage{textcomp}
\usepackage{subfigure}
\usepackage{algorithm}

\usepackage{longtable}

\usepackage{xcolor}

\usepackage{verbatim}

\usepackage{multirow}

\def\BibTeX{{\rm B\kern-.05em{\sc i\kern-.025em b}\kern-.08em
    T\kern-.1667em\lower.7ex\hbox{E}\kern-.125emX}}
\begin{document}

\title{Semantic-aware Speech to Text Transmission with Redundancy Removal}

\author{\IEEEauthorblockN{Tianxiao Han, Qianqian Yang\IEEEauthorrefmark{2}, Zhiguo Shi, Shibo He, Zhaoyang Zhang}
\IEEEauthorblockA{ Zhejiang University, Hangzhou 310007, China\\
Email: \{txhan,qianqianyang20\IEEEauthorrefmark{2},shizg,s18he,ning\_ming\}@zju.edu.cn
}
\thanks{{\thefootnote}{*}This work is partly supported by the SUTD-ZJU IDEA Grant (SUTD-ZJU (VP) 202102), and partly by the Fundamental Research Funds for the Central Universities under Grant 2021FZZX001-20.}
}

\maketitle

\begin{abstract}
Deep learning (DL) based semantic communication methods have been explored for the efficient transmission of images, text, and speech in recent years. In contrast to traditional wireless communication methods that focus on the transmission of abstract symbols, semantic communication approaches attempt to achieve better transmission efficiency by only sending the semantic-related information of the source data. In this paper, we consider semantic-oriented speech to text transmission. We propose a novel end-to-end DL-based transceiver, which includes an attention-based soft alignment module and a redundancy removal module to compress the transmitted data. In particular, the former extracts only the text-related semantic features, and the latter further drops the semantically redundant content, greatly reducing the amount of semantic redundancy compared to existing methods. We also propose a two-stage training scheme, which speeds up the training of the proposed DL model. The simulation results indicate that our proposed method outperforms current methods in terms of the accuracy of the received text and transmission efficiency.  Moreover, the proposed method also has a smaller model size and shorter end-to-end runtime.

\end{abstract}


\section{Introduction}
The continuously increasing demand for communication causes the explosion of wireless data traffic, and places a heavy burden on the current infrastructure of communication systems. Semantic communication is a promising technology for next generation communications because of its great potential of significantly improving transmission efficiency\cite{qin2021semantic_survey}. Unlike traditional communication systems, which focus on transmitting symbols while ignoring semantic content, semantic communication focuses on gathering semantic information from the source and recovering the same semantic information at the receiver. Therefore, concentrated semantic information will be transmitted to the receiver instead of directly mapped bit sequences from the source. By doing so, to transmit the same amount of information, the required resources for semantic communication will be reduced significantly. Moreover, semantic communication has been proved to be more robust than traditional communication systems\cite{qin1}, especially in harsh channel conditions.

The idea of semantic communication has been proposed by Weaver at the beginning of modern communication\cite{shannon1949mathematical}. Following this preliminary work, Carnap and Bar-hillel \cite{carnap1952outline} give an information theoretic definition of semantic information, which is further investigated in \cite{bao2011towards}. A semantic aware data compression method has been proposed in \cite{basu2014preserving} by leveraging a shared knowledge base. However, before the boom of deep learning, there has not been an effective way to actually perform semantic communication of content.

With the emergence of deep learning techniques on image processing and language processing, there have been several works on semantic communications which show the superiority over the traditional methods. A CNN model was presented in \cite{bourtsoulatze2019deep} to enable joint source and channel coding (JSCC) for wireless image transmission, which can recover images under limited bandwidth and low SNR conditions, and achieves efficient image transmission. A deep learning-based semantic communication system has been developed for efficient and robust transmission of text in \cite{qin1}, the deep learning model of which is then further compressed to be able to work on IoT devices \cite{xie2020lite}. \cite{xie2021task} and \cite{xie2021task1} designed semantic communication systems that are capable of multimodal data transmission for tasks, such as visual question answering.

For the transmission of speech, attention-based semantic communication has been developed to recover speech signals at the receiver\cite{Weng2101:Semantic}. A federal learning-based approach has been proposed in \cite{tong2021federated} to further improve the accuracy of recovered speech signals at the receiver. 
For the further semantic purpose of the speech, a speech recognition semantic communication system has been developed in \cite{weng2021semantic}, which reconstructs text transcription of the speech signals at the receiver by transmitting text-related semantic features. However, the connectionist temporal classification (CTC) based approach proposed by \cite{weng2021semantic} encodes each speech spectrum frame into the same amount of transmitted symbols, while ignoring the difference in semantic significance of each frame, which may degrade the transmission efficiency. 

To improve the speech recognition performance and transmission efficiency, we employ an attention-based alignment module to enforce the amount of the semantic features to be transmitted to be close to that of the corresponding text content. All the repeated and semantically irrelevant features are further dropped by a redundancy removal module. Furthermore, we use a trainable semantic decoder at the receiver to transform the semantic feature into the text instead of the greedy decoder used in \cite{weng2021semantic}.

The main contributions of this article can be summarized as follows:
\begin{itemize}
\item We propose a novel speech to text semantic communication approach, which includes an attention-based soft alignment module, which extracts only the text-related semantic features, and a redundancy removal module, which further removes semantically irrelevant features.

\item We apply a two-stage training method, which speeds up the training of the proposed model by training different parts at each stage. 

\item The numerical results validate the effectiveness and efficiency of the proposed method in both the text recognition performance and runtime.

\end{itemize}

The rest of this article is organised as follows. Section \uppercase\expandafter{\romannumeral2} introduces the system model of the considered semantic communication problem and the performance metrics. Section \uppercase\expandafter{\romannumeral3} details the proposed deep learning-based approach semantic communication.  Simulation results are presented in Section \uppercase\expandafter{\romannumeral4} and Section \uppercase\expandafter{\romannumeral5} concludes the paper.

\begin{figure*}[tbp]
\includegraphics[width=\textwidth]{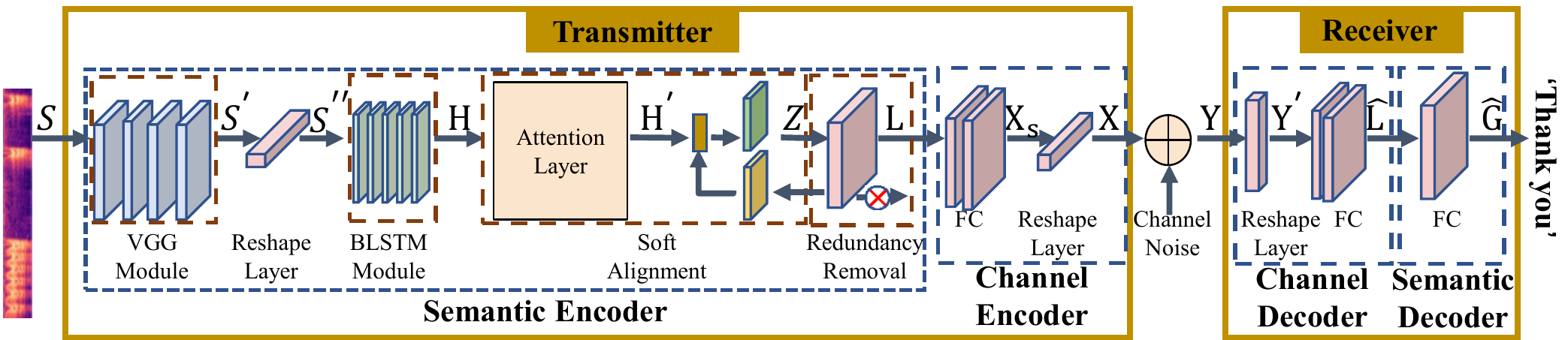} 
\centering 
\caption{The overall architecture of the proposed speech to text semantic communication system}
\label{model}
\end{figure*}

\section{SYSTEM MODEL}
In this section, we present the system model of the considered semantic communication system for speech recognition. We also introduce the metrics to evaluate the performance of the proposed model for speech to text transmission.

\subsection{Transmitter and Receiver}

The considered semantic communication system for speech recognition consists of two parts: the transmitter and the receiver. The speech signal is sampled at 16 kHz and we use a 25 ms Hamming window and 10 ms shift on the input speech signal. And we compute fast Fourier transform(FFT) and filter banks(fbanks)\cite{hermansky1990perceptual } of each Hamming window to get the speech spectrum. The speech spectrum is then input to the transmitter, denoted by $\boldsymbol S=\left[{S_1,S_2,...,S_n }\right]$, where $n$ is the number of frames. The receiver aims to output the corresponding transcription of the input speech, $\boldsymbol G=\left[{G_1,G_2,...,G_m}\right]$, where $G_i\in V$ is the $i^{th}$ word in the transcription, $m$ is the number of words in the transcription, and $V$ denotes the vocabulary, which contains all the possible words in the speech. The mapping between $\boldsymbol S$ and $\boldsymbol G$ is called \textit{alignment} \cite{bahdanau2015neural}, which is a core task for speech recognition.

The transmitter consists of a semantic encoder and a channel encoder. The semantic encoder derives a compact latent semantic representation $\boldsymbol L$ from the input spectrum $\boldsymbol S$. Then, the channel encoder maps $\boldsymbol L$ into symbols, $\boldsymbol X$,to be transmitted over physical channels. The received signal at the receiver is given by \begin{equation} {\boldsymbol Y=\boldsymbol h \ast \boldsymbol X+\boldsymbol w}, \label{channel}  \end{equation} where $\boldsymbol h$ represents the channel coefficients, and $\boldsymbol w\sim\mathcal{CN}(0,\;\sigma^2\mathbf I)$ denotes the independent and identically distributed (i.i.d.) complex Gaussian noise, where $\sigma^2$ is the noise variance, and $\mathbf I$ is the identity matrix.

At the receiver, the received signal, $\boldsymbol Y$, is mapped back into the latent semantic representation $\widehat{\boldsymbol L}$ by a channel decoder, which is then converted into the predicted transcriptions $\widehat{\boldsymbol G}$ by the semantic decoder.

The goal of this paper is to optimize the semantic encoder, the channel encoder, the channel decoder, and the semantic decoder to achieve efficient and robust speech to text transmission.

\subsection{Metric}
We employ the word-error-rate (WER\cite{klakow2002testing}) and the semantic similarity score \cite{qin1} as the performance metrics to evaluate the performance of the considered speech to text transmission. WER is calculated by 
\begin{equation}
WER=\frac{S+D+I}{N},
\label{WER}
\end{equation}
where $S$, $D$, $I$ and $N$ denote the numbers of word substations, word deletions, word insertions, and the number of words in the transcription $\boldsymbol G$, respectively. 

The semantic similarity score to quantify the sentence similarity between the predicted transcription $\widehat{\boldsymbol G}$, and  the original transcription, $\boldsymbol G$, is given by
\begin{equation}\label{similar}
   {\text{similarity}}\left( {\widehat{\boldsymbol G}},{\boldsymbol G} \right) =  \frac{{B \left( { \widehat{\boldsymbol G}} \right) \cdot { B {{\left(\boldsymbol {G} \right)}^T}}}}{{|| {B  {\left( \widehat{\boldsymbol G}\right)}} ||\cdot|| { B {\left(\boldsymbol {G} \right)}} ||}},
\end{equation}
where $ {B(\cdot) }$ represents sentence embedding by a pre-trained text embedding model, Bidirectional Encoder Representations from Transformers (BERT\cite{devlin2018bert}). The sentence similarity score is  a number between 0 and 1, which indicates how similar one sentence is to another,  with 1 representing semantically equivalent and 0 representing not relevant at all.

\section{Proposed Speech to Text Semantic Communication Approach}
In this section, we present the details of the proposed model depicted in Fig. \ref{model}, where the transmitter consists of a semantic encoder and a channel encoder, and the receiver consists of a channel decoder and a semantic decoder. The semantic encoder takes a speech, $\boldsymbol S$, as input and  outputs latent semantic representations, $\boldsymbol L$, where semantic redundancy is reduced by a redundancy removal module. The channel encoder produces a sequence of symbols $\boldsymbol X$ from $\boldsymbol L$ to be transmitted over the physical channel. At the receiver, the received signal $\boldsymbol Y$ is input to the channel decoder to derive the estimated latent semantic representation sequence $\widehat{\boldsymbol L}$. Finally, the semantic decoder decodes $\widehat{\boldsymbol L}$, and outputs the predicted transcription $\widehat{\boldsymbol G}$ from $\widehat{\boldsymbol L}$. We describe each module in detail in the sequel.

\subsection{Semantic Encoder and Decoder}
The semantic encoder has four components, as shown in Fig. \ref{model}, i.e., the VGG module, the BLSTM module, the soft alignment module and the redundancy removal module. 
The input speech spectrum, $\boldsymbol S\in\mathfrak R^{B\times N \times 40 \times 3}$ is acquired by applying 25 ms Hamming window and 10 ms shift on the speech signal and then computing FFT to get 40 fbanks coefficients together with their first- and second-order derivatives \cite{furui1981cepstral}, where $B$ is the batch size, $N$ is the number of frames, 40 is the number of coefficients, and the three channels corresponds to fbanks coefficients and its first- and second-order derivatives, respectively. The sequence of speech spectrums $\boldsymbol S$ is fed into the VGG  module\cite{vgg} to obtain $\boldsymbol S^{\prime}\in\mathfrak R^{B\times \frac{N}{4} \times 10 \times 128}$, that is, 128 feature maps of size $\frac{N}{4} \times 10$ for each input. And the reshape layer concatenates all the 128 channels of $\boldsymbol S^{\prime}$ and output $ \boldsymbol S^{\prime \prime}\in\mathfrak R^{B\times \frac{N}{4} \times 1280}$. Then, $\boldsymbol S^{\prime \prime}$ is fed into a bidirectional Long Short Term Memory (BLSTM) module which generates intermediate features, $\boldsymbol H \in\mathfrak R^{B\times \frac{N}{4 \times 8} \times 512}$, where 8 is the total length reduction\cite{chan2016listen} of BLSTM module. We use downsampling rates of 2,2,2,1,1 for five BLSTM layers in the BLSTM module, respectively, which results in a total downsampling rate of 8.

\begin{figure}[tbp]
\includegraphics[width=0.4\textwidth]{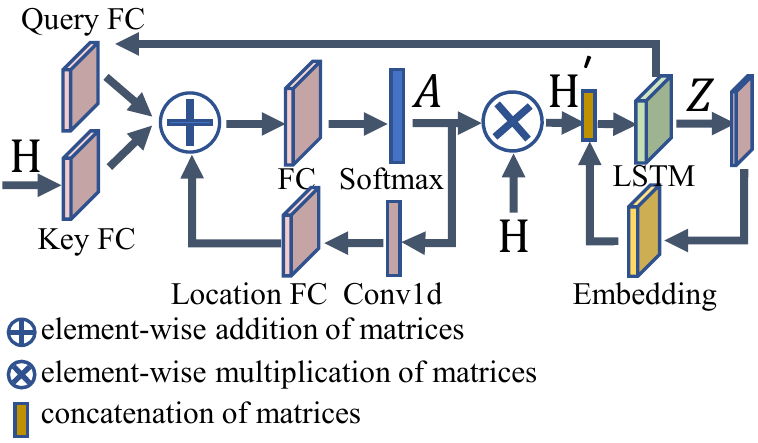} 
\centering 
\caption{The proposed soft alignment module with an attention module and a LSTM layer}
\label{model2}
\end{figure}

\begin{figure}[tbp]
\includegraphics[width=0.45\textwidth]{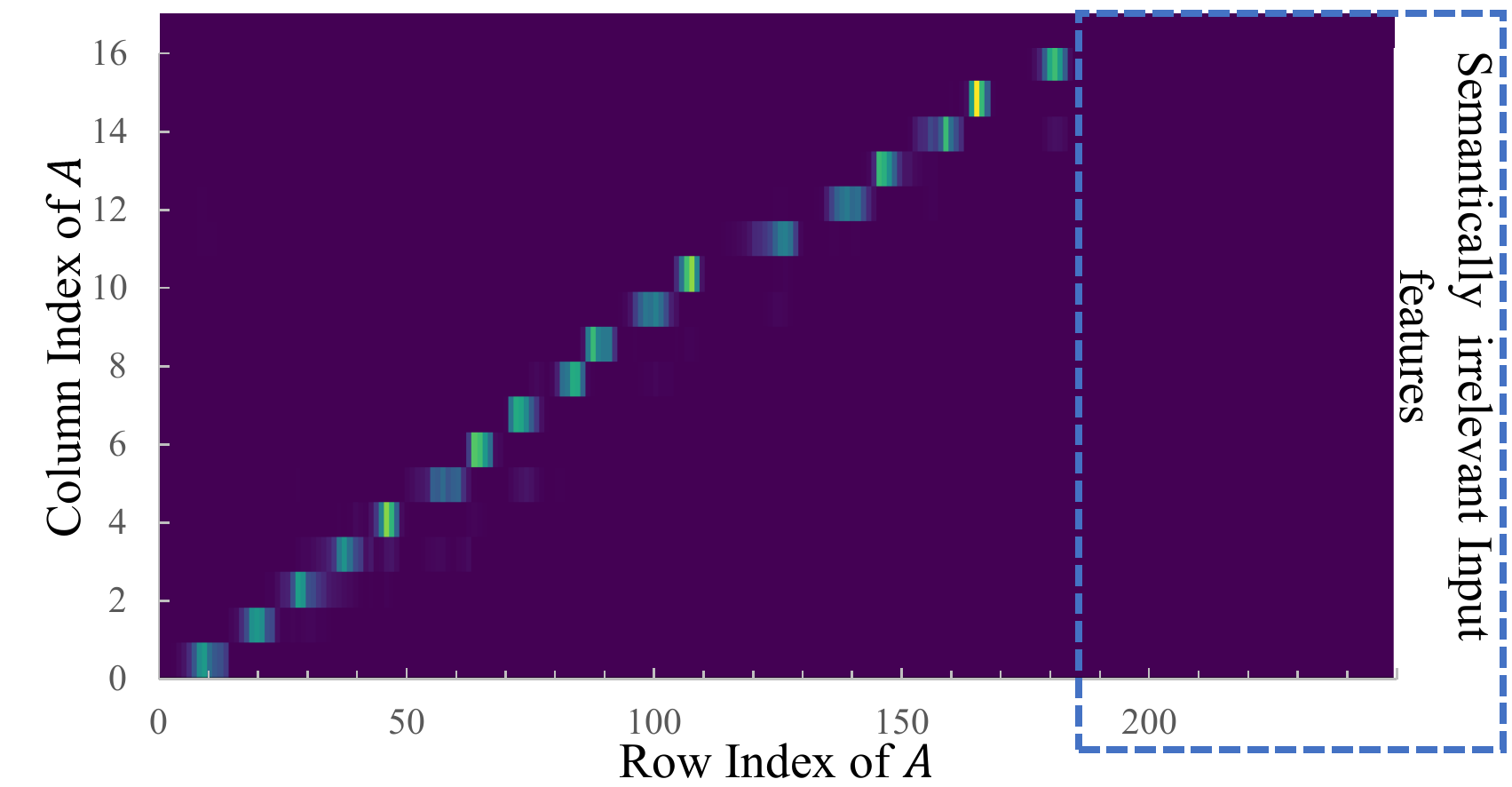} 
\centering 
\caption{Heatmap of attention scores generated by the attention mechanism. Larger values appear yellow, and lower values appear purple.}
\label{atten}
\end{figure}

Next, the intermediate features, $\boldsymbol H$ are fed into the attention-based soft alignment module to extract semantic features with an attention module and a LSTM layer, as shown in Fig. \ref{model2}. The attention mechanism\cite{chorowski2015attention}  is adopted for our soft alignment module to get the alignment of speech with its semantic text. In order to acquire the alignment, we need to compute the weight, also referred to as attention scores, assigned to each element of the input by the query information and the corresponding key information. The query information is derived by passing the hidden states of the LSTM layer through a fully connected layer, and the key information is derived by another fully connected layer with $\boldsymbol H$ as the input. The query and key information are then combined with the feedback information through element-wise addition as shown in Fig. \ref{model2}. Then this combined information is fed into a FC layer and then a softmax layer to get the normalization scores $\boldsymbol A $, which feedback through a 1D convolution layer and a FC layer to derive the location information. Each value in the normalized attention scores $\boldsymbol A \in\mathfrak R^{B \times q \times \frac{N}{4 \times 8}}$ multiplies its corresponding value in $\boldsymbol H$ to get the latent semantic representation, $\boldsymbol H^{\prime} \in\mathfrak R^{B \times q \times 512}$, where $q$ is the variable length of latent semantic representation, which depends on the semantic information the signal is carrying. And $\boldsymbol H^{\prime}$, concatenated with the feedbacked embeddings, is fed into a LSTM layer, which outputs $\boldsymbol Z \in\mathfrak R^{B \times q \times 512}$ as the input to the redundancy removal module. We present one example of the derived attention scores matrix $\boldsymbol A$ in the form of heat map in Fig. \ref{atten}. We can observe that only a few elements of this matrix are with values that are not close to zero. As it is revealed by the numerical results, the proposed soft alignment module leads to better transmission efficiency because the derived attention scores push the transmission resource allocated to semantic significant parts.

The redundancy removal module generates a concentrated latent semantic representation ${\boldsymbol L \in\mathfrak R^{B \times c \times 512}}$, where $c$ is the length of this representation. The detail of the redundancy removal module will be introduced in the sequel. The semantic decoder at the receiver is relatively simple, which consists of a FC layer. It decodes the received text-related semantic features  $\widehat {\boldsymbol L} \in\mathfrak R^{B \times c \times 512}$ into the text transcriptions, $\widehat {\boldsymbol G} \in\mathfrak R^{B \times c \times 15003}$, where $c$ is the length of the transcription.

\subsection{Redundancy Removal Module}
The redundancy removal module consists of a FC layer, which outputs a probability matrix of size ${B \times q \times 15003}$, and each element of the last dimension represents a token in the vocabulary list. The vocabulary list contains the most frequent 15000 words and three special symbols, that is, $EOS$, $UNK$, and $PAD$. The $EOS$ is used at the end of a sentence, the $UNK$ denotes words that are not in the vocabulary list, and $PAD$ is padded to the end of sentences to make the length identical for convenient computing. The redundancy removal module removes these three special tokens and the sequences after the token $EOS$ since they are without any semantic meaning. The experiments reveal that cutting off the sequences after the token $EOS$ saves about 59.4\% of the transmission length, and the removal of $UNK$ and $PAD$ saves approximately 4.5\%. We note that the output $\boldsymbol L$ is also fed into an embedding layer to get an embedding of size ${B \times q \times 512}$ to concatenate with  $\boldsymbol H^{\prime}$ of next time stamp before feeding into the LSTM layer. 

\subsection{Channel Encoder and Decoder}
In the channel encoder, two cascade FC layers map semantic encoder output $\boldsymbol L$ to symbol sequences  $\boldsymbol X_s \in\mathfrak R^{B \times c \times 256}$, which is then reshaped into $\boldsymbol X \in\mathfrak R^{B \times 128c \times 2}$, where the first and second channels are the real parts and imaginary parts of wireless signal, to be transmitted. The received symbol sequences, $\boldsymbol Y \in\mathfrak R^{B \times 128c \times 2}$ at the receiver, are reshaped into $\boldsymbol Y^{\prime} \in\mathfrak R^{B \times c \times 256}$, which are then input to two cascade FC layers to recover text-related semantic features, $\widehat {\boldsymbol L} \in\mathfrak R^{B \times c \times 512}$ to be fed into the semantic decoder.

\subsection{Two-Stage Training}
We use a two-stage training method, which converges faster than training the  network as a whole in an end-to-end manner, as revealed by experiments. In the first stage, we train the semantic encoder and semantic decoder, and ignore the channel encoder and decoder by directly inputting the output of the semantic encoder to the semantic decoder. We use a cross-entropy loss function between the predicted transcription $\widehat {\boldsymbol G}$, and the ground truth for the transcription $\boldsymbol G$ with Adadelta optimizer. Due to the presence of a feedback loop in the soft alignment module, we use the teacher forcing strategy \cite{williams1989learning} to achieve fast convergence. Specifically, at the beginning of the training, as the output of the FC layer is incorrect, we use the true transcriptions instead to speed up the learning of correct alignment.

In the second stage, we freeze the trained semantic encoder and train other parts of the network, including channel encoder, channel decoder, and semantic decoder under physical channel with random SNR. The cross-entropy loss function and Adadelta optimizer are also utilized.

\begin{table}[]
\centering
\footnotesize
\caption{Parameter settings of the proposed semantic communication Network for speech recognition}
\label{table 2}
\begin{tabular}{|c|cc|c|c|}
\hline
                                   & \multicolumn{2}{c|}{Layer Name}                                                    & Parameters                     & Activation \\ \hline
\multirow{17}{5em}{\centering Semantic\\Encoder} & \multicolumn{1}{c|}{\multirow{5}{5em}{\centering VGG\\Module}} & 2$\times$CNN       & 3$\times$3/64            & ReLU       \\ \cline{3-5} 
                                   & \multicolumn{1}{c|}{}                                         & MaxPool            & 2$\times$2 & None       \\ \cline{3-5} 
                                   & \multicolumn{1}{c|}{}                                         & 2$\times$CNN       & 3$\times$3/128               & ReLU       \\ \cline{3-5} 
                                   & \multicolumn{1}{c|}{}                                         & MaxPool            & 2$\times$2                        & None       \\ \cline{3-5} 
                                   & \multicolumn{1}{c|}{}                                         & Reshape            & None                         & None       \\ \cline{2-5} 
                                    & \multicolumn{1}{c|}{\multirow{2}{5em}{\centering BLSTM\\Module}}            & \multirow{2}{*}{5$\times$BLSTM} & \multirow{2}{5em}{\centering 512}         & \multirow{2}{*}{Tanh} \\
                                  & \multicolumn{1}{c|}{}                                         &                                &                              &                       \\ \cline{2-5} 
                                  
                                   & \multicolumn{1}{c|}{\multirow{7}{5em}{\centering Soft\\Alignment\\Module}}        & Query FC          & 300                          & Tanh       \\ \cline{3-5} 
                                   & \multicolumn{1}{c|}{}                                         & Key FC          & 300                          & Tanh       \\ \cline{3-5} 
                                   & \multicolumn{1}{c|}{}                                         & Conv1d     & 201/10/100                 & None       \\ \cline{3-5} 
                                   & \multicolumn{1}{c|}{}                                         & FC  & 300                       & Tanh       \\ \cline{3-5} 
                                   & \multicolumn{1}{c|}{}                                         & FC   & 1                            & Tanh       \\ \cline{3-5} 
                                   & \multicolumn{1}{c|}{}                                         & LSTM               & 512                      & Tanh       \\ \cline{3-5} 
                                   & \multicolumn{1}{c|}{}                                         & Embedding    & 15003/512                    & None       \\ \cline{2-5} 
                                   & \multicolumn{1}{c|}{\multirow{3}{5em}{\centering Redandancy\\Removal\\Module}} & \multirow{3}{*}{FC}            & \multirow{3}{*}{150003}      & \multirow{3}{*}{None} \\
                                  & \multicolumn{1}{c|}{}                                           &                                &                              &                       \\
                                  & \multicolumn{1}{c|}{}                                           &                                &                              &                       \\ \hline

\multirow{3}{5em}{\centering Channel\\Encoder}   & \multicolumn{2}{c|}{\multirow{2}{*}{FC}}                                           & 512                          & ReLU       \\ \cline{4-5} 
                                   & \multicolumn{2}{c|}{}                                                              & 256                          & None       \\ \cline{2-5} 
                                   & \multicolumn{2}{c|}{Reshape}                                                       & None                          & None       \\ \hline
\multirow{3}{5em}{\centering Channel\\Decoder}   & \multicolumn{2}{c|}{Reshape}                                                       & None                          & None       \\ \cline{2-5} 
                                   & \multicolumn{2}{c|}{\multirow{2}{*}{FC}}                                           & 256                          & Relu       \\ \cline{4-5} 
                                   & \multicolumn{2}{c|}{}                                                              & 512                          & None       \\ \hline
\multirow{2}{5em}{\centering Semantic\\Decoder}  & \multicolumn{2}{c|}{\multirow{2}{*}{FC}}                                           & \multirow{2}{*}{15003}       & \multirow{2}{*}{None} \\
                                   & \multicolumn{2}{c|}{}                                                              &                              &                       \\ \hline

\end{tabular}
\end{table}

\section{Numerical Results}

\begin{algorithm}[htb]
    \caption{Testing algorithm of the proposed method.}
    \label{testing algorithm}
    \begin{algorithmic}[1]   
        \STATE \textbf{Input:}  Speech signals and transcriptions $\boldsymbol G$ from dataset, fading channel $\boldsymbol h$, noise $\boldsymbol w$.
        
        \STATE Set fading channel $\boldsymbol h$ = Rayleigh or AWGN
	    \FOR{each SNR value}
        \STATE Generate Gaussian noise $\boldsymbol W$ under the SNR value.
        \STATE Generate spectrum sequences $\boldsymbol S $ from input speech signals.
        \STATE Output $\boldsymbol L$ from $\boldsymbol S $ by the semantic encoder.
        \STATE Output $\boldsymbol X$ from $\boldsymbol L^{\prime}$ by the channel encoder.
        \STATE Transmit $\boldsymbol X$ and receive $\boldsymbol Y$ via (\ref{channel}).
        \STATE Output $\widehat{\boldsymbol L}$ from $\boldsymbol Y$ by the channel decoder.
        \STATE Output $\widehat {\boldsymbol G}$  from $\widehat{\boldsymbol L^{\prime}}$ by the semantic decoder.
        \ENDFOR
    	\STATE \textbf{Output:} compare $\widehat{\boldsymbol G}$ and $\boldsymbol G$ and compute WER scores and sentence similarity via \ref{WER} 
    \end{algorithmic}
\end{algorithm}

In this section, we compare the proposed method's performance to the existing two other deep learning-based semantic communication systems for speech recognition task. We consider the AWGN and Rayleigh channels for the evaluation. We use the Librispeech dataset\cite{panayotov2015librispeech} for training and testing, which is a speech to text library based on public domain audio books. We use the existing semantic communication approach by \cite{weng2021semantic}, referred to as DeepSC-SR as the benchmark approach to compare to. We also combine the proposed semantic encoder, which transfers the speech signal to semantic text, and the semantic communication approach proposed in \cite{qin1}, which transmits and recovers text at the receiver. This approach is referred to as SE-DeepSC, which is also used as a benchmark to compare with. We use the proposed semantic encoder and decoder, while ignoring the noisy channel as well as channel encoder and decoder, which provides the upper-bound performance. The detailed setting of our proposed network is shown in table \ref{table 2}. We note that all these four approaches are trained and tested with the same datasets. And the test algorithm is described in Algorithm \ref{testing algorithm}.

\begin{figure}[tbp]
\includegraphics[width=0.4\textwidth]{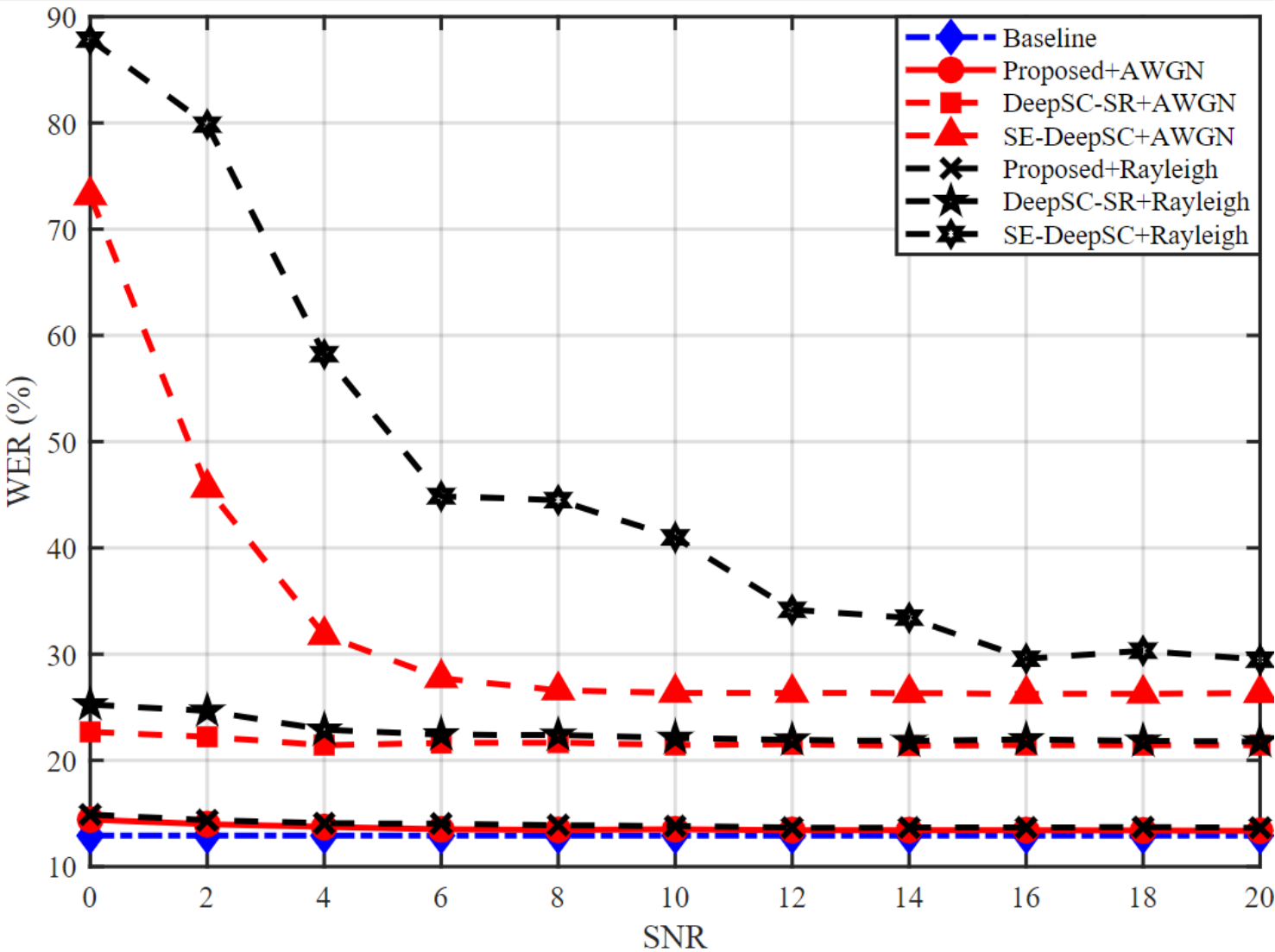}
\centering 
\caption{WER score versus SNR for different approaches.}  
\label{WER result}  
\end{figure}

\begin{figure}[tbp]
\includegraphics[width=0.4\textwidth]{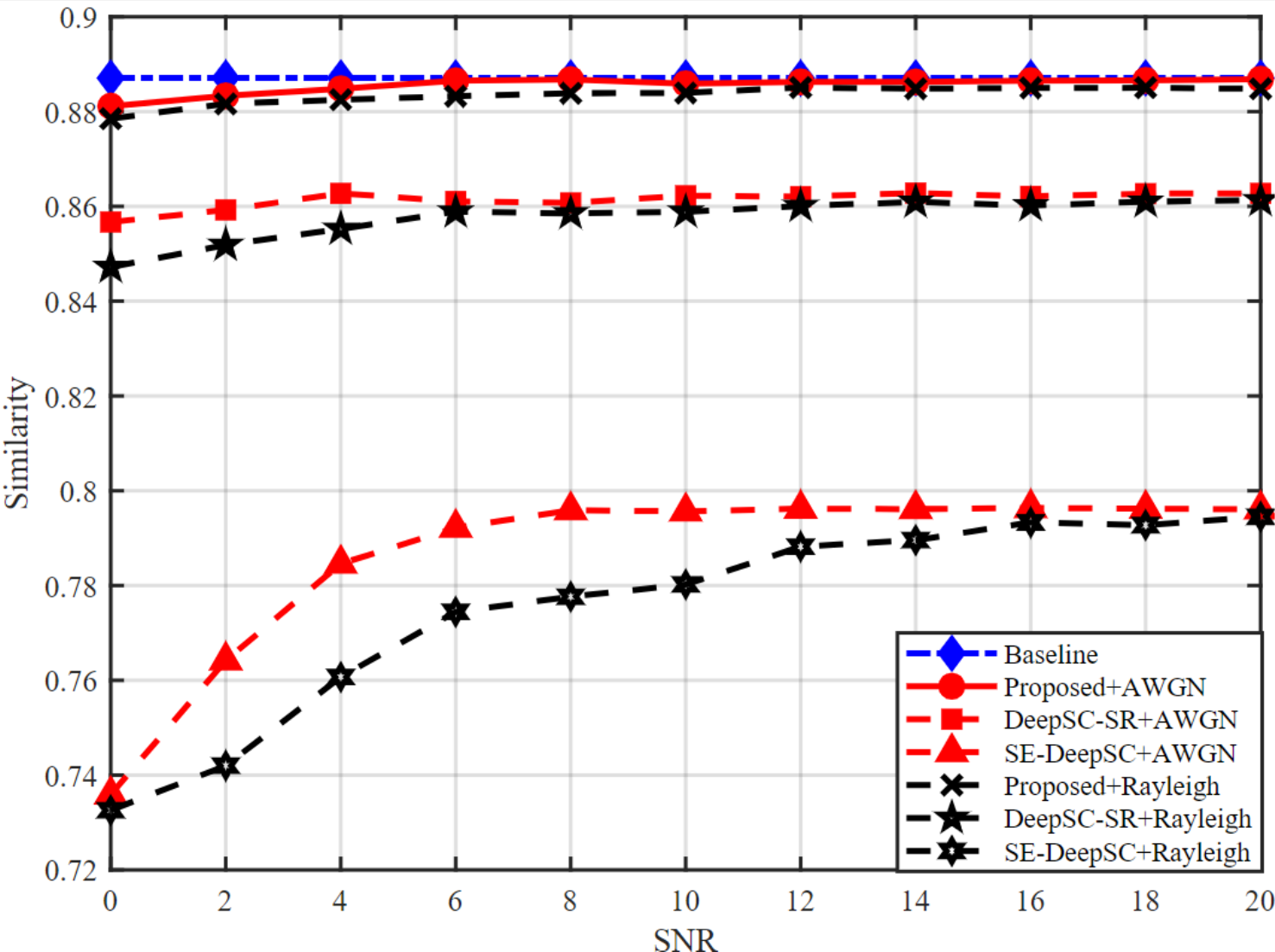}
\centering 
\caption{Sentence similarity score versus SNR for different approaches.}  
\label{sim result}  
\end{figure}

\subsection{WER and Semantic Similarity of the Proposed Model}
The performance comparison of different approaches in terms of WER is presented in Fig. \ref{WER result}. We can see that the proposed method significantly outperforms the other two methods under both channel conditions. Moreover, our proposed method performs steadily and closely to the baseline while the performance of the two benchmarks is poor under a low SNR regime. We note that different from the original result shown in \cite{weng2021semantic} that we get the WER of around 20\% instead of 40\% for DeepSC-SR, which may benefit from the use of 40 dimension fbank features with their first and second derivatives.

The performance comparison of different approaches in terms of sentence similarity score are shown in Fig. \ref{sim result}. We observe that our proposed method obtains higher sentence similarity scores than the other methods under the considered channel conditions, especially in a low SNR regime. It can be observed from both figures that SE-DeepSC performs not as well as the other two counterparts. This may be due to the semantic errors in the output of the semantic encoder, which is neglected in the transmission and reconstruction, and hence, exists in the final output.

\begin{table}[]
\caption{}
\label{table_number}
\begin{tabular}{|c|c|c|}
\hline
                                                                     & Proposed model        & DeepSC-SR             \\ \hline
\multirow{2}{15em}{The length of each transmitted symbol vector}        & \multirow{2}{*}{256}  & \multirow{2}{*}{40}   \\
                                                                     &                       &                       \\ \hline
\multirow{2}{15em}{The average numbers of transmitted symbols per sentence} & \multirow{2}{*}{5156} & \multirow{2}{*}{7143} \\
                                                                     &                       &                       \\ \hline
\end{tabular}
\end{table}

\begin{table*}[htbp]
  \centering
  \caption{Examples of input speech signals and their transcriptions before and after the redundancy removal module.}
  \label{tablelong}
  \begin{tabular}{|c|p{0.8\linewidth}|}
  \hline
    Example 1  & The number of speech spectrum frames: 497 \\
    & Transcription: (Length:34) $UNK$ HE SAID YOU KNOW WHERE $UNK$ $EOS$ $EOS$ $EOS$ $EOS$ $EOS$ $EOS$ $EOS$ $EOS$ $EOS$ $EOS$ $EOS$ $EOS$ $EOS$ $EOS$ $EOS$ $EOS$ $EOS$ $EOS$ $EOS$ $EOS$ $EOS$ $EOS$ $EOS$ $EOS$ $EOS$ $EOS$ $EOS$ \\ saved: 85.3\%
& Transcription after the redundancy removal module: (Length:5) HE SAID YOU KNOW WHERE \\ \hline
Example 2 & The number of speech spectrum frames: 9845 \\
& Transcription: (Length:72) I HAVE DRAWN UP A LIST OF ALL THE PEOPLE WHO OUGHT TO GIVE US A PRESENT AND I SHALL TELL THEM WHAT THEY OUGHT TO GIVE IT WON'T BE MY FAULT IF I DON'T GET IT $EOS$ $EOS$ $EOS$ $EOS$ $EOS$ $EOS$ OF ALL THE PEOPLE WHO OUGHT TO GIVE US A PRESENT AND I SHALL TELL THEM WHAT THEY OUGHT TO GIVE IT WON'T BE MY FAULT IF I DON'T GET IT $EOS$ $EOS$ $EOS$ $EOS$ $EOS$ $EOS$ OF ALL THE PEOPLE WHO OUGHT TO GIVE US A PRESENT AND I SHALL TELL THEM WHAT THEY OUGHT TO GIVE IT WON'T BE MY FAULT IF \\ saved: 45.8\%
& Transcription after the redundancy removal module: (Length:39) I HAVE DRAWN UP A LIST OF ALL THE PEOPLE WHO OUGHT TO GIVE US A PRESENT AND I SHALL TELL THEM WHAT THEY OUGHT TO GIVE IT WON'T BE MY FAULT IF I DON'T GET IT \\ \hline
\end{tabular}
\end{table*}


\subsection{Transmission Efficiency}

We also present the length of the transmitted symbol vector and the average number of the transmitted symbols per sentence on the same testing dataset in Table. \ref{table_number}. We can see from this table that although our proposed model has much longer symbol vectors, i.e., a larger dimension of the output of the channel encoder, the average number of transmitted symbols per sentence by the proposed model is still much smaller than that by DeepSC-SR. The reason may be that DeepSC-SR encodes every single speech spectrum frame into the same amount of transmitted symbols, while the proposed method ignores the redundant content, and only sends the text-related semantic information.

Two examples of the speech signals and their corresponding transcriptions as well as the transcription after redundancy removal are shown in Table. \ref{tablelong}. Both of the examples show that the number of the speech spectrum frames is much larger than the length of the transcription, which implies many of the frames are semantically irrelevant. We also observe that the redundancy removal module significantly reduces the length of the transcription while preserving the semantic meaning.

\begin{table}[htbp]
\caption{The average sentence processing runtime versus various schemes and their model size. }
\label{table_size}
\centering
\begin{tabular}{ |c| c |c |c|} 
\hline
& Parameters &  Model Size &  Runtime (single thread) \\
\hline
Proposed  & 66582925&  255MB &  0.097 s \\
\hline
DeepSC-SR & 83724534 &  320MB &  0.120 s \\
\hline
SE-DeepSC & 69405509 &  349.2MB & 0.246 s \\
\hline
\end{tabular}
\end{table}

\subsection{Model Size and Runtime}
We compare the computational complexities and memory cost of the proposed model and the two benchmark approaches.  All the experiments run on the same server with a single NVIDIA GeForce RTX 3090 GPU. We present the average runtime per sentence and the model sizes in Table. \ref{table_size}. We can see that our proposed model has a smaller model size than the other two approaches, as well as a shorter running time.

\section{Conclusions}
In this paper, we propose a novel semantic-aware speech to text transmission approach with soft alignment module and redundancy removal module. The attention-based soft alignment module is utilized to get the alignment of the input speech and its latent semantic representation. The redundancy removal module drops the semantically irrelevant features to obtain a more compact semantic representation. Simulation results show that our proposed method adapts well to diverse channels and improves accuracy significantly. We also propose a two-stage training approach that reduces the training complexity. We showed numerically the proposed approach outperforms the existing approach in terms of the speech recognition accuracy and the transmission efficiency, benefited from the soft alignment module and the redundancy removal module. The proposed model also has a smaller model size and shorter runtime than the counterpart approach. One possible future direction is to extend the proposed model for the real-time speech to text transmission of the steaming speech signals.

\bibliographystyle{IEEEtran}
\bibliography{reference.bib}

\begin{thebibliography}{10}
\providecommand{\url}[1]{#1}
\csname url@samestyle\endcsname
\providecommand{\newblock}{\relax}
\providecommand{\bibinfo}[2]{#2}
\providecommand{\BIBentrySTDinterwordspacing}{\spaceskip=0pt\relax}
\providecommand{\BIBentryALTinterwordstretchfactor}{4}
\providecommand{\BIBentryALTinterwordspacing}{\spaceskip=\fontdimen2\font plus
\BIBentryALTinterwordstretchfactor\fontdimen3\font minus
  \fontdimen4\font\relax}
\providecommand{\BIBforeignlanguage}[2]{{%
\expandafter\ifx\csname l@#1\endcsname\relax
\typeout{** WARNING: IEEEtran.bst: No hyphenation pattern has been}%
\typeout{** loaded for the language `#1'. Using the pattern for}%
\typeout{** the default language instead.}%
\else
\language=\csname l@#1\endcsname
\fi
#2}}
\providecommand{\BIBdecl}{\relax}
\BIBdecl

\bibitem{qin2021semantic_survey}
Z.~Qin, X.~Tao, J.~Lu, and G.~Y. Li, ``Semantic communications: Principles and
  challenges,'' \emph{arXiv preprint arXiv:2201.01389}, 2021.

\bibitem{qin1}
H.~Xie, Z.~Qin, G.~Y. Li, and B.-H. Juang, ``Deep learning enabled semantic
  communication systems,'' \emph{IEEE Trans. Signal Process.}, pp. 2663--2675,
  Apr. 2021.

\bibitem{shannon1949mathematical}
C.~Shannon and W.~Weaver, ``The mathematical theory of communication. univ. of
  illinois press, urbana.'' \emph{The mathematical theory of communication.
  Univ. of Illinois Press, Urbana.}, 1949.

\bibitem{carnap1952outline}
R.~Carnap, Y.~Bar-Hillel \emph{et~al.}, ``An outline of a theory of semantic
  information,'' 1952.

\bibitem{bao2011towards}
J.~Bao, P.~Basu, M.~Dean, C.~Partridge, A.~Swami, W.~Leland, and J.~A. Hendler,
  ``Towards a theory of semantic communication,'' in \emph{2011 IEEE Network
  Science Workshop}.\hskip 1em plus 0.5em minus 0.4em\relax IEEE, 2011, pp.
  110--117.

\bibitem{basu2014preserving}
P.~Basu, J.~Bao, M.~Dean, and J.~Hendler, ``Preserving quality of information
  by using semantic relationships,'' \emph{Pervasive and Mobile Computing},
  vol.~11, pp. 188--202, 2014.

\bibitem{bourtsoulatze2019deep}
E.~Bourtsoulatze, D.~B. Kurka, and D.~G{\"u}nd{\"u}z, ``Deep joint
  source-channel coding for wireless image transmission,'' \emph{IEEE Trans.
  Cogn. Commun. Netw.}, vol.~5, no.~3, pp. 567--579, Sept. 2019.

\bibitem{xie2020lite}
H.~Xie and Z.~Qin, ``A lite distributed semantic communication system for
  {I}nternet of {T}hings,'' \emph{IEEE J. Sel. Areas Commun.}, vol.~39, no.~1,
  pp. 142--153, Jan. 2021.

\bibitem{xie2021task}
H.~Xie, Z.~Qin, X.~Tao, and K.~B. Letaief, ``Task-oriented multi-user semantic
  communications,'' \emph{arXiv preprint arXiv:2112.10255}, 2021.

\bibitem{xie2021task1}
H.~Xie, Z.~Qin, and G.~Y. Li, ``Task-oriented multi-user semantic
  communications for vqa task,'' \emph{IEEE Wireless Communications Letters},
  2021.

\bibitem{Weng2101:Semantic}
Z.~Weng and Z.~Qin, ``Semantic communication systems for speech transmission,''
  \emph{IEEE J. Sel. Areas Commun.}, Apr. 2021.

\bibitem{tong2021federated}
H.~Tong, Z.~Yang, S.~Wang, Y.~Hu, O.~Semiari, W.~Saad, and C.~Yin, ``Federated
  learning for audio semantic communication,'' \emph{Frontiers in
  Communications and Networks}, vol.~2, 2021.

\bibitem{weng2021semantic}
Z.~Weng, Z.~Qin, and G.~Y. Li, ``Semantic communications for speech
  recognition,'' \emph{arXiv preprint arXiv:2107.11190}, 2021.

\bibitem{hermansky1990perceptual}
H.~Hermansky, ``Perceptual linear predictive (plp) analysis of speech,''
  \emph{the Journal of the Acoustical Society of America}, vol.~87, no.~4, pp.
  1738--1752, 1990.

\bibitem{bahdanau2015neural}
D.~Bahdanau, K.~H. Cho, and Y.~Bengio, ``Neural machine translation by jointly
  learning to align and translate,'' in \emph{3rd International Conference on
  Learning Representations, ICLR 2015}, 2015.

\bibitem{klakow2002testing}
D.~Klakow and J.~Peters, ``Testing the correlation of word error rate and
  perplexity,'' \emph{Speech Communication}, vol.~38, no. 1-2, pp. 19--28,
  2002.

\bibitem{devlin2018bert}
J.~Devlin, M.-W. Chang, K.~Lee, and K.~Toutanova, ``Bert: Pre-training of deep
  bidirectional transformers for language understanding,'' \emph{arXiv preprint
  arXiv:1810.04805}, 2018.

\bibitem{furui1981cepstral}
S.~Furui, ``Cepstral analysis technique for automatic speaker verification,''
  \emph{IEEE Transactions on Acoustics, Speech, and Signal Processing},
  vol.~29, no.~2, pp. 254--272, 1981.

\bibitem{vgg}
K.~Simonyan and A.~Zisserman, ``Very deep convolutional networks for
  large-scale image recognition,'' in \emph{International Conference on
  Learning Representations}, May 2015.

\bibitem{chan2016listen}
W.~Chan, N.~Jaitly, Q.~Le, and O.~Vinyals, ``Listen, attend and spell: A neural
  network for large vocabulary conversational speech recognition,'' in
  \emph{2016 IEEE International Conference on Acoustics, Speech and Signal
  Processing (ICASSP)}.\hskip 1em plus 0.5em minus 0.4em\relax IEEE, 2016, pp.
  4960--4964.

\bibitem{chorowski2015attention}
J.~Chorowski, D.~Bahdanau, D.~Serdyuk, K.~Cho, and Y.~Bengio, ``Attention-based
  models for speech recognition,'' \emph{arXiv preprint arXiv:1506.07503},
  2015.

\bibitem{williams1989learning}
R.~J. Williams and D.~Zipser, ``A learning algorithm for continually running
  fully recurrent neural networks,'' \emph{Neural computation}, vol.~1, no.~2,
  pp. 270--280, 1989.

\bibitem{panayotov2015librispeech}
V.~Panayotov, G.~Chen, D.~Povey, and S.~Khudanpur, ``Librispeech: an asr corpus
  based on public domain audio books,'' in \emph{2015 IEEE international
  conference on acoustics, speech and signal processing (ICASSP)}.\hskip 1em
  plus 0.5em minus 0.4em\relax IEEE, 2015, pp. 5206--5210.

\end{thebibliography}

\end{document}